\documentclass{article}
\usepackage[utf8]{inputenc}
\usepackage[square,numbers]{natbib}
  \usepackage{lineno}
\usepackage{graphicx}
\usepackage{authblk}
\usepackage[usenames,dvipsnames]{color}
\usepackage{indentfirst}
\usepackage{amssymb}
\usepackage{amsmath}

\begin{document}

\title{Force generation by a cylindrical cell under stationary osmolytes synthesis}

\author[1]{Weiyuan Kong}
\author[1]{Antonio Mosciatti Jofré}
\author[2]{Manon Quiros}
\author[3]{Marie-Béatrice Bogeat-Triboulot}
\author[2]{Evelyne Kolb}
\author[1]{Etienne Couturier}
\affil[1]{Laboratoire Matière et Systèmes Complexes, Université Paris Diderot CNRS UMR 7057, 10 Rue Alice Domont et Léonie Ducquet, 75205 Paris Cedex 13, France}
\affil[2]{PMMH, CNRS, ESPCI Paris, Université PSL, Sorbonne Université, Université de Paris, F-75005, Paris, France}
\affil[3]{Université de Lorraine, AgroParisTech, INRAE, UMR Silva, 54000 Nancy, France}

\date{\today}

\begin{abstract}

Turgor is the driving force of plant growth, making possible for roots to overcome soil resistance or for stems to counteract gravity. Maintaining a constant growth rate while avoiding the cell content dilution, which would progressively stop the inward water flux, imposes the production or import of osmolytes in proportion to the increase of volume. We coin this phenomenon stationary osmoregulation. The article explores the quantitative consequences of this hypothesis on the interaction of a cylindrical cell growing axially against an obstacle.

An instantaneous axial compression of a pressurized cylindrical cell generates a force and a pressure jump which both decrease toward a lower value once water has flowed out of the cell to reach the water potential equilibrium. In a first part, the article derives analytical formula for these force and over-pressure both before and after relaxation. In a second part, we describe how the coupling of the Lockhart's growth law with the stationary osmoregulation hypothesis predicts a transient slowdown in growth due to contact before a re-acceleration in growth. We finally compare these predictions with the output of an elastic growth model which ignores the osmotic origin of growth: models only match in the early phase of contact for high stiffness obstacle.

\end{abstract}
\maketitle

\section{Introduction}\label{Intro}
\begin{figure}[htbp]
\begin{center}
\includegraphics[scale=0.7]{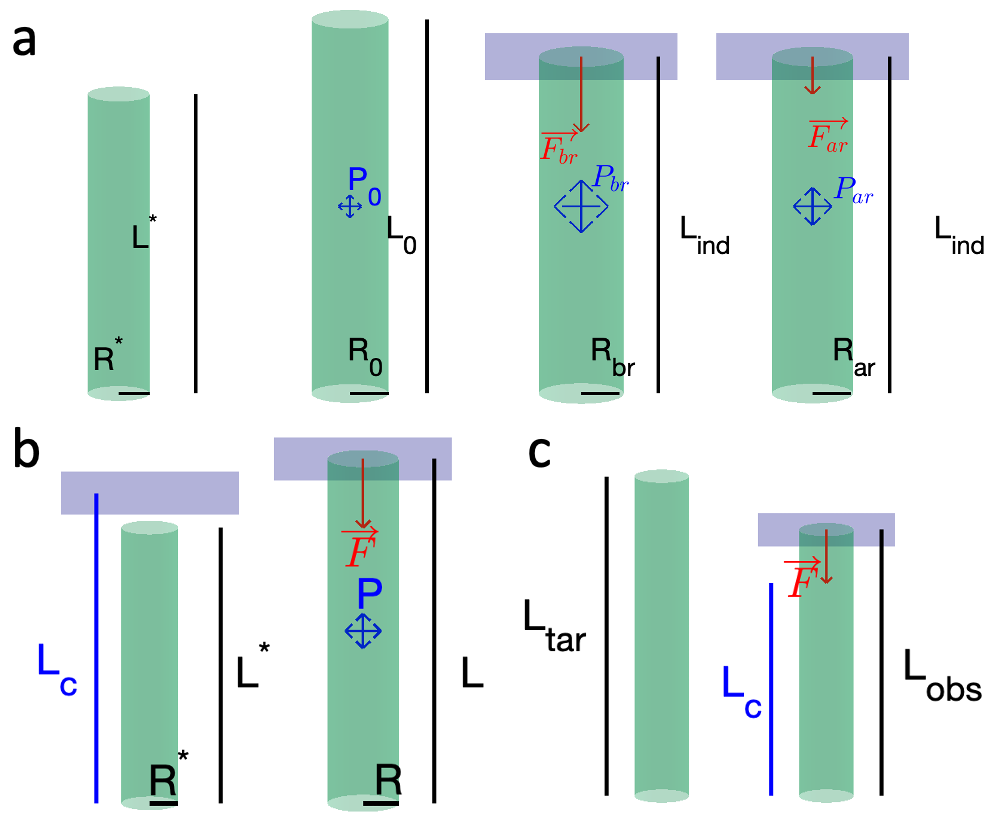}
\caption{Schema of the theoretical scenarii. a. Inflation and compression scenario. From left to right: Plasmolyzed cell, turgid cell, compressed turgid cell before relaxation and compressed turgid cell after relaxation. $R^*$ (resp. $L^*$) are the plasmolyzed radius (resp. the plasmolyzed length). $R_0$ (resp. $L_0$) are the turgid radius (resp. the turgid length) before compression. $L_{ind}$ is the length of the compressed cell.When the cell undergoes compression from the obstacle (in violet), $R_{br}$ and $P_{br}$  are the turgid radius and turgor pressure before relaxation, while $R_{ar}$ and $P_{ar}$ are the turgid radius and turgor after relaxation.  b. Growth scenario against an obstacle. From left to right: Schema of a plasmolyzed cell and of the obstacle (in violet), schema of the turgid cell once the growth has pushed the obstacle. $R^*$ (resp. $L^*$) are the plasmolyzed radius (resp. the  plasmolysed length). $L_c$ is the length of the turgid cell at contact. $R$ (resp. $L$) are the turgid radius (resp. the turgid length) during the contact and $F$ is the force exerted by the obstacle. c. Schema of the morphoelastic model. $L_{tar}$ stands for the target length and $L_{obs}$ stands for the observed length, once the growth has pushed the obstacle.}
\label{Experiences}
\end{center} 
\end{figure}

 Plant growth requires water fluxes that are generated by gradients of water potential between the growing cells and the water source. The water potential gradient is tightly regulated by the growing cell through both osmoregulation and cell wall relaxation, a viscous process, which acts on the volume of the plasmolyzed cell. These regulations modify the cell internal pressure, called turgor pressure \citep{boyer1985water}. The maintenance of turgor pressure and thus of osmotic pressure requires osmolyte synthesis or import to compensate for the increase of the cell volume. We coin this process "stationary osmoregulation". The generated driving force is sufficient to counteract the resistance of the surrounding fluids (air, water, tissue or soil). 
 Before touch-induced regulations \citep{braam2005touch} have time to be effective or osmolyte production rate to be altered, the pressure pattern within a growing organ encountering a rigid obstacle will be theorically modified by the resistance opposed by the obstacle, the pressure rising in the most impeded parts. On a longer time scale, hard obstacles are counteracted by the active modification of the cell turgor, through the increase of osmotic pressure, as previously shown in roots \citep{greacen_physics_1972},  \citep{atwell_physiological-responses_1988}.\\
Turgor can in certain conditions positively regulate cell wall growth through the Lockhart's law, which tells that the growth rate allowed by cell wall relaxation is proportional to the pressure above a threshold \citep{proseus2000turgor}. Lockhart’s model can even be a quantitative tool to predict how the force caused by obstacle modifies the growth dynamics \citep{minc2009mechanical}, \citep{quiros2022plant}. Meanwhile, the coupling between Lockhart's law and osmolyte production has been mainly studied in the context of growth rate oscillation for pollen tubes \citep{hill2012osmotic}, \citep{dumais2021mechanics} rather than in the context of force generation.\\
Cylindrical organs, omnipresent in the vegetal world, range from unicellular cylindrical internodes of \textit{Characeae}, to multicellular plant stems and roots. The present note is focused on a single cylindrical cell presenting a surface extending homogeneously in the axial direction, which is coined monoaxial diffuse growth \citep{cosgrove2018diffuse}. We quantify how an obstacle will modify the cell wall stress pattern and the cell internal pressure as well as the growth dynamics. More generally our model can be applied to other "walled" cells of organisms other than plants, such as the fungus \textit{Phycomyces blackeslaneus} or the Gramm positive bacteria  \textit{Bacillus subtilis}: these organisms present a cell wall and a positive regulation of growth by turgor similar to the one observed in plants \citep{ortega_comparison_1991}, \citep{rojas2018regulation}.\\
The mechanical analysis of giant cellular internode was traditionally considered as broadly applicable to the multicellular cylindrical organs (stems, hypocotyls ...) \citep{kutschera1995tissue}. However patterns of cell wall stress in multicellular cylindrical organs can be more complex: recent studies have suggested that the maximal stress direction in the epidermis of hypocotyls is axial, contrarily to single cells where it is circumferential \citep{robinson2018global}, \citep{verger2018tension}.
Another difference between a single cylindrical cell and a cylindrical organ lays on water potential gradients: for a single cell, the water potential gradient necessary to move water into the cell during growth is small compared to the internal pressure \citep{zhu_enlargement_1992} while for multicellular organs, in general the water potential gradient cannot be neglected in front of the internal pressure \citep{nonami1993direct}. Nevertheless the present results modeled for a single cell can serve as a valuable guide to interpret the effect of an obstacle on the growth of cylindrical multicellular organs with similar stress patterns.\\
Two theoretical scenarii are explored theoretically and solved analytically: the first scenario consists in compressing the cell to impose instantaneously a smaller length than the natural length (Figure \ref{Experiences}a) and the second scenario consists in the progressive loading of the cell by its proper axial growth (Figure \ref{Experiences}b). Without an increase of the osmolyte content through import or internal production, the water inflow in the growing cell that increases the protoplast volume would dilute the osmolytes, decrease the osmotic pressure and thus decrease turgor. Our model quantifies how the force generated by a stationary osmolyte synthesis/import and the elastic force (cell wall, obstacle) retroact on growth through the Lockhart’s law. Parameters can be directly estimated from pressure probe literature \citep{proseus2000turgor}. The growth scenario is an opportunity to compare our physiological model to a phenomenological approach, coined morpho-elasticity, based on an analogy with metal thermo-elasticity \citep{goriely2011morphoelasticity}.

\section{Determination of the apparent stiffness by a compression scenario}\label{section1}
\begin{figure}[htbp]
\begin{center}
\includegraphics[scale=0.4]{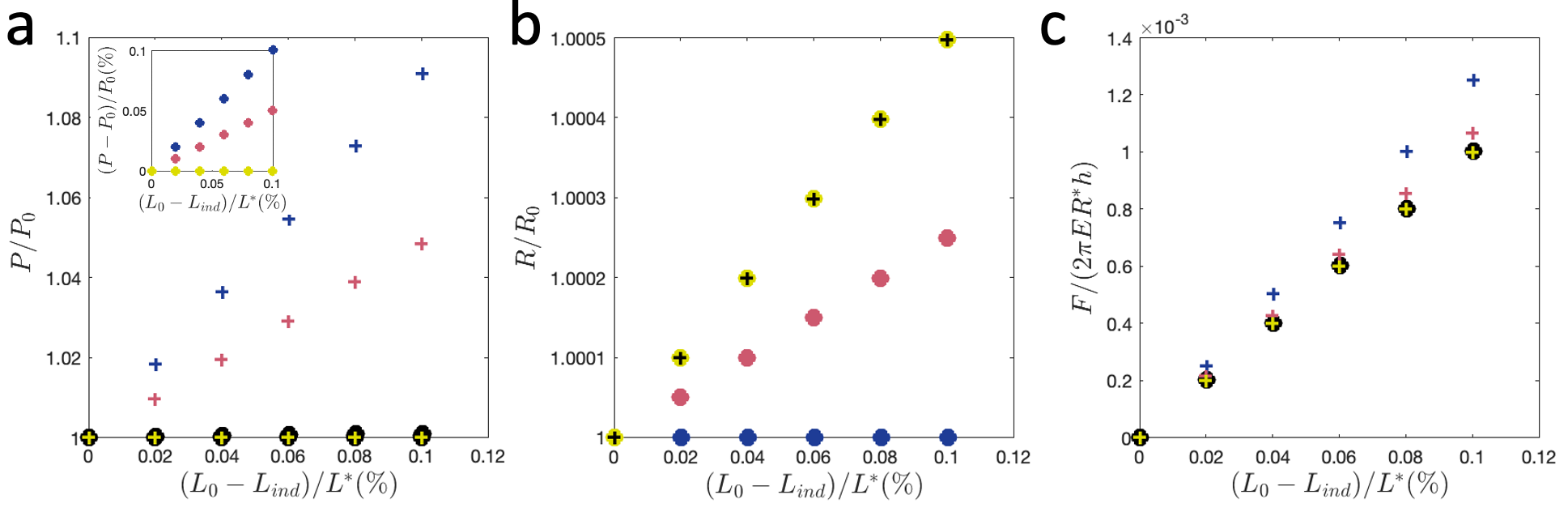}
\caption{Relationship between parameters in the compression scenario. The symbols $+$ stand for the data before relaxation and filled circles $\bullet$ for the data after relaxation. The color code of symbols tells the Poisson ratio $\nu$ (blue $\nu=0$, red $\nu=0.25$, yellow $\nu=0.5$). Black color is used when the variable is independent of $\nu$. a. Normalized pressure $P/P_0$, $P_0$ being the turgor before compression. The pressure increment after relaxation, $(P_{ar}-P_0)/P_0$,  which is of the same order as the longitudinal strain, is represented in the top left insert.
b. Normalized radius $R/R_0$.
c. Normalized force $F/(2\pi E R^* h)$ vs relative longitudinal strain due to the compression $(L_0-L_{ind})/L^*$.}
\label{Experience manona}
\end{center} 
\end{figure}
We aim to measure the apparent stiffness of a cylindrical cell of turgor $P_0$ by compressing the cell to a given length and measuring the resulting force. It corresponds experimentally to a compression of the cell with a force sensor of infinite stiffness. We start with a cylindrical cell without turgor, whose plasmolyzed length and radius are $L^*$ and $R^*$ respectively and with a cell wall thickness $h\ll R^*$. The cell wall is mechanically characterized by a Young's modulus $E$ and a Poisson ratio $\nu$. The cell is inflated at turgor $P_0$ and the resulting turgid length and turgid radius are $L_0$ and $R_0$ respectively. According to the elastic theory of thin shells (valid for wall thickness much smaller than cylinder radius and for small strains), $h$ stays constant (at first order) and the cell wall longitudinal and circumferential (hoop) stresses before compression are given respectively by $\sigma_{LL,0}=P_0R_0/(2h)$
, $\sigma_{\Theta\Theta,0}=P_0R_0/h$ and the radius and the length are provided by the two classical formula for cylindrical cells \citep{timoshenko1959theory}:
\begin{equation*}
R_0=\frac{R^*}{1-P_0R^*(2-\nu)/(2Eh)},
\end{equation*}
\begin{equation*}
L_0=\frac{L^*(1-(1+\nu)P_0R^*/(2Eh))}{1-P_0R^*(2-\nu)/(2Eh)},
\end{equation*}

We suppose the cell to be at water potential equilibrium which gives the following osmotic (molar) content \citep{jensen2022soft}: $N_i=\pi L_0 R_0^2(\Pi_{ext}+P_0)/(\mathcal{R}T),$ $\Pi_{ext}$ being the external osmotic pressure and $\mathcal{R}$ the perfect gas constant. We latter suppose $\Pi_{ext}=0$. Considering the high membrane hydraulic conductivity 
we suppose $N_i$, the number of moles of osmolytes to stay constant 
until the water potential equilibrium.\\
The compression scenario consists in imposing instantaneously a new length, $L_{ind}$ to the cell; as water has no time to move out of the cell, the volume is at first the same as before the compression, which imposes the radius before relaxation to be:
$R_{br}=R_0\sqrt{L_0/L_{ind}}.$ This statement supposes that there is no barreling of the cell when compressed (see Annex \ref{non_linearities}). Longitudinal and circumferential stresses before relaxation read:
$\sigma_{LL,br}=P_{br}R_{br}/(2h)-F_{br}/(2\pi R_{br} h)$, $\sigma_{\Theta\Theta,br}=P_{br}R_{br}/h$.
For small elastic deformation, the mechanical equilibrium reads:
\begin{equation}
\left(\begin{array}{c}
 L_{ind}/L^*-1  \\
 R_{br}/R^*-1
\end{array} \right)=\left(\begin{array}{cc}
 1& -\nu  \\
-\nu& 1
\end{array} \right)\left(\begin{array}{c}
P_{br}R_{br}/(2Eh) -F_{br}/(2\pi E R_{br} h) \\
P_{br}R_{br}/(Eh)
\end{array} \right)
\label{mechanica_equilibrium_br}
\end{equation}
 $P_{br}$, being the pressure before relaxation and $F_{br}$, the force before the relaxation. \\
 Consistently with the linear elasticity hypothesis, the second order term can be neglected, which gives the following simple formula (See Annex \ref{Force and pressure before relaxation}):
\begin{eqnarray*}
R_{br}&=&R_0+R^*\frac{(L_0-L_{ind})}{2L^*},\\
P_{br}&=&P_{0}+\frac{Eh}{R^*}\frac{1-2\nu}{2(1-\nu^2)}\frac{L_0-L_{ind}}{L^*}\\
F_{br}&=&(2\pi ER^*h)\left(\frac{5}{4}-\nu\right) \frac{L_0-L_{ind}}{L^*(1-\nu^2)}.
\end{eqnarray*}
Figure \ref{Experience manona} (with symbols $+$) shows how these parameters linearly increase with the longitudinal strain, the slope depending in particular on the cell wall Poisson ratio $\nu$ for the evolution of turgor $P_{br}$ and force $F_{br}$. The apparent stiffness $k_{br}$ of the cell before relaxation, defined from $F_{br}=k_{br} \, (L_0-L_{ind})$ reads: $$k_{br}=\left(\frac{5/4-\nu}{1-\nu^2}\right)\frac{2\pi ER^*h}{L^*}.$$

Once water has moved out of the cell and the water potential equilibrium is reached, the pressure and radius after relaxation ($P_{ar}$ and $R_{ar}$) are linked by: 
$$P_{ar}=\mathcal{R} T N_i/(\pi R_{ar}^2 L_{ind}).$$
Substituting the expression for $P_{ar}$ in the mechanical equilibrium provides:
\begin{equation}
\left(\begin{array}{c}
 L_{ind}/L^*-1  \\
 R_{ar}/R^*-1
\end{array} \right)=\left(\begin{array}{cc}
 1& -\nu  \\
-\nu& 1
\end{array} \right)\left(\begin{array}{c}
\mathcal{R} T N_i/(2E h\pi R_{ar} L_{ind}) -F_{ar}/(2\pi E R_{ar} h) \\
\mathcal{R} T N_i/(E h\pi R_{ar} L_{ind})
\end{array} \right)
\label{mech_eq_osm_eq}
\end{equation}
Neglecting the second order term yields the radius, pressure and force after relaxation  (See Annex \ref{Force and pressure after relaxation}  and Figure \ref{Experience manona} (with symbols $\bullet$)):
\begin{eqnarray}
\label{after_relaxation}
R_{ar}&=&R_0+\nu R^* \frac{(L_0-L_{ind})}{L^*}\nonumber\\
P_{ar}&=&P_0\left(1+(1-2\nu)\frac{(L_0-L_{ind})}{L^*}\right)\\
F_{ar}&=&(2\pi ER^*h)\frac{(L_0-L_{ind})}{L^*}.\nonumber
\end{eqnarray}
The apparent stiffness after relaxation reads: 
\begin{equation}
k_{ar}=\frac{2\pi ER^*h}{L^*}.\label{stiff_ar}
\end{equation}
For a one centimeter long \textit{Chara corallina} internodal cell, the calculation of the stiffness gives a value of $k_{ar}=3141\,N\,m^{-1}$ (See Annex \ref{Analytical solution} for the parameters).
A similar stiffness was derived in \citep{tuson2012measuring} by physical argument.\\
The formula \ref{after_relaxation} for $P_{ar}$ tells that the pressure in an hemipermeable cell impeded by an obstacle will rise quicker under the water inflow. The water potential equilibrium will be reached for a smaller volume increment and at a higher pressure than without the obstacle. Though the water potential equilibrium is not a valid hypothesis in general for growing tissues \citep{martre1999measurement}, the encounter with a rigid obstacle should modify the pressure pattern in a growing tissue before inducing other responses: the pressure will rise in the most deformed parts tightening the water potential gradient and altering the water fluxes toward these regions.\\

The forces before and after relaxation depend solely on the Young's modulus, the thickness of the cell wall, the plasmolyzed radius and the Poisson ratio, not on the initial turgor.
$\Delta P\, = P_{br}-P_{ar}$, (resp. $\Delta F\, = F_{br}-F_{ar}$) the difference of pressure (resp. force) before and after relaxation, are zero for $\nu=0.5$ and maximal for $\nu=0$:
\begin{eqnarray*}
\Delta P&=&(1-2\nu)\left(\frac{Eh}{2R^*(1-\nu^2)}-P_0\right)\frac{L_0-L_{ind}}{L^*},\\
\Delta F&=&(2\pi ER^*h)\frac{\left(\frac{1}{2}-\nu\right)^2}{1-\nu^2} \frac{L_0-L_{ind}}{L^*}.
\end{eqnarray*}
The linear elasticity hypothesis implies:
$\frac{P_0R^*}{Eh}\ll 1$. Though not visible with the scales of main Figure \ref{Experience manona}a, the insert shows that the pressure after relaxation is higher than the initial turgor for $\nu <0.5$. However the jump between the initial turgor and the pressure after relaxation (ie. $P_{ar}-P_0$) being small compared to $P_{br}-P_{0}$, $\Delta P$ can be approximated by $P_{br}-P_{0}$. 

\section{Growth interaction with an obstacle under stationary osmolyte synthesis}\label{section2}
Growth can be introduced in the model by adding two equations, one for the production or import of osmolytes and one for the cell wall relaxation. Other processes such as the change of mechanical properties of the cell wall due to maturation will not be considered herein. The growth of giant internodal cell is one-dimensional and exponential. We coin "stationary osmoregulation" an exponential osmolyte production, which compensate the dilution due to the exponential growth: $N_i=N_i(0)\exp(\gamma t)$, $\gamma$ being the osmolyte production rate. For giant internodal cells, the water potential gradient necessary to drive water into the cell can be neglected in front of the absolute value of cell turgor and osmotic pressure \citep{zhu_enlargement_1992}; $P$, the turgor pressure satisfies the following relation:
\begin{equation}
P=(\mathcal{R}TN_i)/(\pi L R^2)-\Pi_{ext}.
\label{Pressure}
\end{equation}
For matter of simplicity we suppose the external osmotic pressure $\Pi_{ext}$ to be $0$. As the cell is always supposed to be at the water potential equilibrium, we call cell stiffness the stiffness of the cell after relaxation (Formula \ref{stiff_ar}):
$$k_{cell}=k_{ar}.$$
The obstacle can be described as a spring of stiffness $k$. At the time of contact set at $t=0$, $L_c$ is the turgid length of the cell and $L_{c}^*$ is the plasmolyzed length. At $t>0$, $L$ (respectively $L^*$) stands for the turgid (resp. plasmolyzed) length and the force is $F=k(L-L_c)$. The stresses read:
$\sigma_{LL}=PR/(2h)-F/(2\pi R h)$
and $\sigma_{\Theta\Theta}=PR/h.$ Substituting the pressure expression provides:
\begin{equation}
\sigma_{LL}=(\mathcal{R}TN_i)/(2\pi h R L)-F/(2\pi R h),
\label{stress_axial_growth}
\end{equation}
\begin{equation}
\sigma_{\Theta\Theta}=(\mathcal{R}TN_i)/(\pi h R L).
\label{stress_radial_growth}
\end{equation}
The mechanical equilibrium is supposed to be always satisfied: solving (\ref{Pressure}, \ref{stress_axial_growth}, \ref{stress_radial_growth}) while neglecting second order terms following linear elasticity hypothesis (see Annex \ref{osmotic equilibrium and mechanical}) provides the longitudinal and circumferential (hoop) strains in the cell wall:
\begin{equation}
\epsilon_{LL}=\frac{(1-2\nu)\mathcal{R} T N_i/(2Eh\pi R^* L^*)-k(L^*-L_c)/(2\pi E R^* h)}{1+k(2L^*-L_c)/(2\pi E R^* h)},
\end{equation}
\begin{equation}
\epsilon_{\Theta\Theta}=(1-\nu^2)\mathcal{R} T N_i/(Eh\pi R^* L^*)-\nu\epsilon_{LL}.
\end{equation}
The longitudinal stress reads:
\begin{equation}
\sigma_{LL}/E=\epsilon_{LL}+\nu\mathcal{R} T N_i/(Eh\pi R^* L^*).
\label{formule_stress_suggestion}
\end{equation}
In some cylindrical organs at a late stage of development (stems, internodes ...), the cylindrical organ grows solely in length. The growth allowed by the relaxation dynamics of the cell wall will be introduced by an equation on the dynamics of the plasmolyzed length:
\begin{equation}
\frac{d L^*}{d t}=f(L^*,t)
\label{Lockhart_0}
\end{equation}
The function $f$ could correspond to different scenarii. In the simplest one, there is no retroaction of the force on relaxation ($f$ is proportional to the rest length) but this scenario does not describe the phenomenology of the growth slowdown after contact observed for several experimental models: root \citep{bizet_3d_2016}, pollen tube \citep{burri2018feeling}, root hair \citep{pereira2022mechanical}, Gramm positive bacteria \citep{tuson2012measuring}, fission yeast \citep{minc2009mechanical}. The next step of complexity is the Lockhart's law which tells that growth rate is proportional to the pressure above a threshold $f_P=L^*m_P(P-Y_P)_+$ \citep{lockhart_analysis_1965}. In \textit{characea} internodes, the linearity of Lockhart's law above the yield threshold has been shown for up to $120\%$ of the initial turgor during long lasting measurements (more than 5 hours) (See \citep{proseus2000turgor}, Figure 5 and 7). Lockhart's law can also be formulated in cell wall strains $f_\epsilon=L^*m_\epsilon(\epsilon_{LL}-Y_\epsilon)_+$ \citep{boudon2015computational} or stresses $f_\sigma=L^*m_\sigma(\sigma_{LL}-Y_\sigma)_+$ \citep{huang2012modelling}. As a compression tends to slightly increase the turgor after relaxation (See Section \ref{section1}), the traditional pressure formulation of the Lockhart's law would predict a slight acceleration of the growth during the contact, which is contrary to most experimental observations. Herein $f_\sigma$, the stress formulation will be used as it has been recently proven that it can quantitatively predict how the maize root growth slows down when encountering an axial rigid obstacle \citep{quiros2022plant}. Formally speaking, the stress formulation is equivalent to a Bingham-type rheology for the cell wall, where the extensibility in stress ($m_\sigma$ in $Pa^{-1}.s^{-1}$) is the inverse of a plastic viscosity. After contact, the equation for the stress formulation reads:
\begin{equation}
\frac{d L^*}{d t}=m_\sigma E\left(\frac{(1-2\nu)\mathcal{R} T N_i/(2Eh\pi R^* )-kL^*(L^*-L_c)/(2\pi E R^* h)}{1+k(2L^*-L_c)/(2\pi E R^* h)}+\frac{\nu\mathcal{R} T N_i}{\pi E R^*h }-\frac{Y_\sigma L^*}{E}\right)_+.
\label{Lockhart_stress}
\end{equation}
$\gamma$, the osmolyte production rate is chosen such as the pressure remains constant before the contact. As the cell grows solely in length, the volume increase rate equals the length increase rate before the contact:
\begin{equation*}
\frac{dN_i/dt}{N_i}=\frac{dL^*/dt}{L^*},
\end{equation*}
which provides $\gamma$:
\begin{equation}
\gamma\approx \frac{f(L_{c}^*,0)}{L_{c}^*}.
\end{equation}
To generalize and simplify the formulation of equations, we introduced the following non-dimensional variables:

\begin{equation}
\hat{L}^*=\frac{L^*}{L_c^*},\ \hat{P}=\frac{\mathcal{R}TN_{i}}{2\pi R^* hL_{c}^*E},\ \hat{P}_c=\frac{\mathcal{R}TN_{i}(0)}{2\pi R^* hL_{c}^*E},\ \hat{Y}_\sigma=\frac{Y_\sigma}{E},
\label{nondim1}
\end{equation}
\begin{equation}
\hat{k}=\frac{k}{k_{cell}},\ \hat{\gamma}=  \left(\hat{P_c}-\hat{Y}_\sigma\right)_+,\ \hat{t}=m_\sigma E t
\label{nondim2}
\end{equation}

The cell wall being mechanically characterized  by its Young's modulus $E$, a straightforward way to obtain non-dimensional (ND) cell wall stresses is to divide them  by $E$. Before contact, the longitudinal stress $\sigma_{LL}$ is directly proportional to the turgor $P$ through a geometrical factor $\frac{R}{2h}$. Incorporating the formula (\ref{Pressure}) for turgor into $\frac{\sigma_{LL}}{E}$ provides the ND longitudinal elastic stress $\hat{P}$ before contact. In the same way, $\hat{P}_c$ and $\hat{Y}_\sigma$ are introduced and correspond to the ND longitudinal elastic stress at contact and to the ND yield threshold. Thus $\hat{\gamma}$ is the ND increment above the threshold for an elastic stress formulation. From the Lockhart's extensibility $m_\sigma$, the parameter $1/(m_\sigma E)$ has the dimension of a time and is naturally introduced to define the ND time $\hat{t}$.

For \textit{Chara corallina} for example, the characteristic time $1/(m_\sigma E)$ is $960 s$ (See Annex \ref{Analytical solution}). Linear elasticity supposes small strains which correspond to a normalized pressure $\hat{P}_c$ of less than a few percent. The normalized Lockhart's threshold $\hat{Y}_\sigma$ is also less than a few percent because the cell is supposed to grow before the contact occurs ($Y<P_c$).\\

These non-dimensional variables can be used to rewrite (\ref{Lockhart_stress}) (See Annex \ref{Analytical solution}):
\begin{equation}
\frac{d \hat{L}^*}{d \hat{t}}=\hat{f}(\hat{L}^*,\hat{t})
\label{Lockhart_stress_adim}
\end{equation}
with:
\begin{equation}
\hat{f}(\hat{L}^*,\hat{t})=\left(\frac{(1-2\nu)\hat{P}_c\exp(\hat{\gamma}\hat{t})-\hat{k}\hat{L}^*(\hat{L}^*-1-(1-2\nu)\hat{P}_c)}{1+\hat{k}(2\hat{L}^*-1-(1-2\nu)\hat{P}_c)}+2\nu\hat{P}_c\exp(\hat{\gamma}\hat{t})-\hat{Y}_\sigma\hat{L}^*\right)_+.
\end{equation}

The equation was numerically solved with Matlab function ode23s (See \ref{Analytical solution} for the parameters of the ODE which corresponds to an internodal cell \textit{Chara corallina} whose length at contact with the obstacle is $1\,cm$.
\begin{figure}[htbp]
\begin{center}
\includegraphics[scale=0.9]{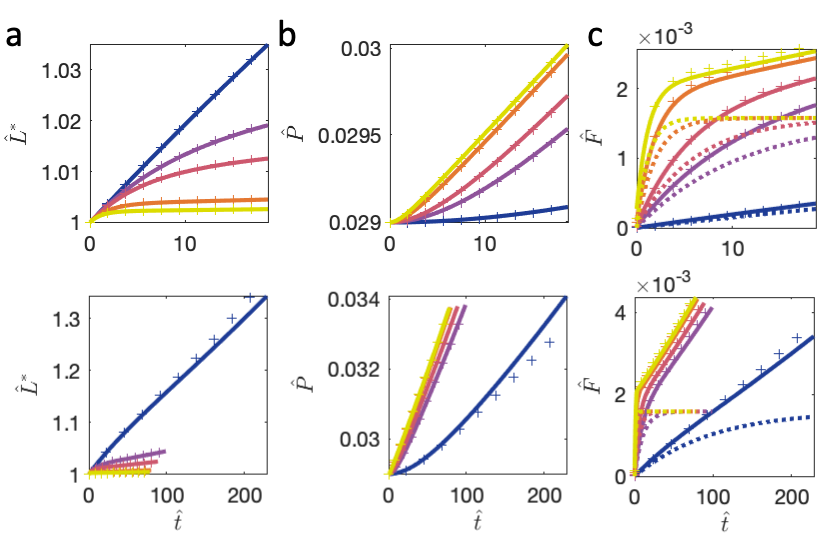}
\caption{(a) Normalized plasmolyzed length, (b) normalized pressure and (c) normalized force vs normalized time for different normalized stiffness $\hat{k}=1/100$ (blue), $\hat{k}=1/10$ (violet), $\hat{k}=1/5$ (dark red), $\hat{k}=1$ (orange), $\hat{k}=5$ (yellow). Other parameters are detailed in the annex (\ref{Analytical solution}). Plain lines, $+$ symbol, dotted line stand for the numerical solution, the analytical solution and the morphoelastic solution respectively. The upper plot is a zoom on the initial behavior ($5$ first hours) whereas the lower plot is integrated until one of the strains equals $3\%$.
The morphoelastic solution is not represented on panel a and b because neither the plasmolyzed length nor the turgor pressure are relevant in morphoelasticity.}
\label{Simu_analytique}
\end{center} 
\end{figure}
The initial growth dynamics can be studied by solving analytically the linearized problem associated with (\ref{Lockhart_stress_adim}) (See Annex \ref{Analytical solution}) which provides a combination of two exponential functions:
\begin{equation}
\hat{L}^*(\hat{t})\approx 1+\hat{\beta}\left(\exp\left(\hat{\gamma} \hat{t}\right)-1\right)+\Delta\hat{L}^*\left(1-\exp\left(-\hat{\gamma}_i \hat{t}\right)\right),
\label{Analytic}
\end{equation}
and:
\begin{equation*}
\hat{\gamma}_i= \frac{\hat{k}}{1+\hat{k}}+\hat{Y}_\sigma,\ 
\hat{\beta}=\frac{\hat{P_c}\left(1+2\nu\hat{k}\right)}{(1+\hat{k})\hat{P_c}+\hat{k}},\ \Delta\hat{L}^*=(1-\hat{\beta})\frac{\hat{\gamma}}{\hat{\gamma}_i}.
\end{equation*}

It corresponds to a normalized turgid length:
\begin{equation}
\hat{L}(\hat{t})\approx \hat{L}^*(\hat{t})(1+\epsilon_{LL}(\hat{t})),
\end{equation}
which after substitution reads:
\begin{equation*}
\hat{L}(\hat{t})\approx 1+(1-2\nu)\hat{P}_c+\frac{\hat{P}_c(\exp(\hat{\gamma}\hat{t})-1)(1-2\nu+\hat{k})}{(1+\hat{k})((1+\hat{k})\hat{P_c}+\hat{k})}+\Delta\hat{L}\left(1-\exp\left(-\hat{\gamma}_i \hat{t}\right)\right)
\label{Turgid_length}
\end{equation*}
with :
\begin{equation*}
 \Delta\hat{L}=\frac{1-\hat{\beta}}{1+\hat{k}}\frac{\hat{\gamma}}{\hat{\gamma}_i}.
\end{equation*}
It corresponds to a normalized force:
\begin{equation}
\hat{F}(\hat{t})\approx \hat{k}(\hat{L}-\hat{L}_c)
\label{Force_model0}
\end{equation}
which reads after substitution of $\hat{L}$
and $\hat{L}_c$:
\begin{equation}
\hat{F}(\hat{t})\approx \frac{\hat{k}}{1+\hat{k}}\left(\frac{\hat{P}_c(\exp(\hat{\gamma}\hat{t})-1)(1-2\nu+\hat{k})}{((1+\hat{k})\hat{P_c}+\hat{k})}+(1-\hat{\beta})\frac{\hat{\gamma}}{\hat{\gamma}_i}\left(1-\exp\left(-\hat{\gamma}_i \hat{t}\right)\right)\right).
\label{Force_model}
\end{equation}
In the limit $\hat{k} \gg 1$, and after the initial relaxation phase ($\hat{\gamma}\hat{t}>3$), the normalized force is independent of the stiffness and follows the normalized pressure (See Figure \ref{Simu_analytique} lower panel):
\begin{equation}
\hat{F}(\hat{t})\approx \hat{P}_c(\exp(\hat{\gamma}\hat{t})-1).
\label{asymptoticForce_model}
\end{equation}
For \textit{Chara corallina}, the characteristic time of the positive exponential is $1/\gamma=213\ h$ while the characteristic time of the negative exponential ranges from $1/\gamma_i=21\ h$ for very low stiffness to $1/\gamma_i=16\ min$ for high stiffness.
For various organs (maize roots \citep{frensch_rapid_1995}, \textit{Characeae} internodes \citep{proseus2000turgor} $\cdots$), $(P-Y)/Y$ lays between $1/5$ and $1/10$, it implies that $\gamma/\gamma_i$ is in general less than $1/5$. The linearized solution for the plasmolyzed length (equation (\ref{Analytic})) is the sum of a positive exponential (characteristic time $1/\hat{\gamma}$) and a negative exponential with a smaller characteristic time, $1/\hat{\gamma_i}$ : on short time scale (Figure \ref{Simu_analytique}(a) upper plot) the solution relaxes toward a plateau according to the negative exponential while on longer time it follows the positive exponential (Figure \ref{Simu_analytique}(a) lower plot) dictated by the osmolyte production rate. For high stiffnesses ($\beta\approx \frac{\hat{P_c}\left(2\nu\right)}{(1+\hat{P_c})}$ which is second order term according to the linear elasticity hypothesis)   the transient plateau is more pronounced than for low stiffnesses ($\beta\approx 1$) as the respective weight of the two exponential functions in the sum shifts from almost $0$ to $1$ (Figure \ref{Simu_analytique}(a)). Analytical and numerical solutions match very well. Regardless of stiffness, the plasmolyzed length keeps diverging exponentially for high $t$ (See Annex \ref{Asymptotic behavior}):
\begin{equation}
L^*\sim_{t\to\infty} \frac{2\nu N_i(t)}{\pi (R^*)^2(P_0+Eh/R^*)}.
\label{infty}
\end{equation}
At $t=0$, the turgid cell is under positive longitudinal elastic strain, while for high $t$ (equation (\ref{infty})) longitudinal strains are strongly negative $\epsilon_{LL}=-50\%$, outside the hypothesis of linear elasticity. In order to ensure that the numerical model output remains in the linear elasticity hypothesis, the maximal integration time (Figure \ref{Simu_analytique}) was chosen such that both strains remain small ($<3\%$).\\
Due to the constant osmolyte production rate and the slow-down of growth rate after contact, the pressure keeps increasing with time (Figure \ref{Simu_analytique}(b)). For an obstacle of low stiffness, ($k\leqslant k_{cell}/10$), the stationary osmoregulation is more effective and the pressure remains fairly constant in the early phase of the interaction with the obstacle (Figure \ref{Simu_analytique} (b), blue curve) while for an obstacle of high stiffness ($k\geqslant k_{cell}$), the pressure increases from the beginning of the interaction (Figure \ref{Simu_analytique} (b), yellow curve). For a $1\ cm$ \textit{Chara corallina} internodal cell, an increase of  $1$ bar is reached in approximately $5$ hours for an obstacle stiffness of $31.42\ N.m^{-1}$ ($k_{cell}/100$) and in $90\ min$ for a stiffness of $3142\ Nm^{-1}$ ($k_{cell}$). Lockhart's growth law has been monitored on this time scale, for $5$ hours (see reference \citep{proseus2000turgor}, Figure 5). In this timeframe, pressure remains below $120\%$ of its initial value (Figure \ref{Simu_analytique} (b)), that is in the linear regime of the Lockhart's law (according to \citep{proseus2000turgor}, Figure 7). 
Besides non-linear effects, some adaptations to external stresses (compression or tension) such as microfibrill reorientation, have been observed on \textit{Characeae} but on longer time scale (four days) \citep{gertel_cell-growth_1977}. The force dynamics presents two successive regimes: a first phase of steep rise (Figure \ref{Simu_analytique} (c) upper panel) followed by a second phase of weaker increase resulting from the turgor pressure increase dynamics (Figure \ref{Simu_analytique} (c) lower panel). For a $1\,cm$ \textit{Chara corallina} internodal cell and for high stiffness obstacle, the transition occurs for a force of a few tenth of $mN$ after a time lag of a tenth of minutes. During the whole simulation represented in Figure \ref{Simu_analytique}, the force remains lower than the Euler criterion force, preventing instabilities such as buckling or barreling (See Annex \ref{non_linearities}).

For multicellular cylindrical organs (roots and stems), the cell wall Young's modulus and the cell wall thickness vary among cells layers and are not uniquely defined. The well-defined variable is the global organ stiffness. Then, to ease the comparison of single-cell with multicellular cylindrical organs, 
non-dimensionalized parameters are expressed with the cell stiffness 
and with the pressure formulation of the Lockhart's parameter (as obtained from the pressure probe litterature): 
\begin{equation}
\hat{t}=m_P P_{\infty} t,\  \hat{P}=\frac{P}{P_{\infty}},\  \hat{Y}_\sigma=\frac{Y_P}{P_{\infty}},\ 
\hat{\gamma}=  \frac{\left(P-Y_P\right)_+}{P_{\infty}},\ \hat{\gamma}_i=\frac{\hat{k}}{1+\hat{k}}+ \frac{Y_P}{P_{\infty}},\ P_{\infty}=\frac{k_{cell}L^*_c}{\pi (R^*)^2}.
\label{parametre_raideur}
\end{equation}

If we could approximate the \textit{Zea mays} root growth zone 
by a single cylindrical cell at water potential equilibrium and diffuse growth (See Annex \ref{Analytical solution}),  $1/(m_P P_{\infty})=50s$  would be twenty times smaller than for \textit{Chara corallina}, and the characteristic time of the positive exponential 
would be $1/\gamma=3\ h$ while the characteristic time of the negative exponential would range from $1/\gamma_i=33\ min$ for very low stiffness to $1/\gamma_i=97\ s$ for high stiffness.

\section{Comparison with an elastic growth model}\label{section3}
The growing cell is now modeled as a growing spring in series with one spring for the obstacle. This morphoelastic model is similar to a model developed for the growth of bacteria in a gel \citep{tuson2012measuring}; we present it here for comparison with the physiological model developed above. The  analytical solution of  the compression scenario provides the  cell spring stiffness. Growth is phenomenologically introduced by increasing the rest cell length  (coined target length $L_{tar}$). The target length increase rate and the negative force retro-acting on growth can be incorporated in the model thanks to two phenomenological constants ($c_1$ and $c_2$ (See Annex \ref{Elastic growth model})) calibrated with the analytical solution. Elastic model totally ignores water fluxes; it is thus not easy to incorporate the stationary osmoregulation hypothesis.\\
The elastic growth model (See Annex \ref{Elastic growth model}) provides an expression for the temporal evolution of the non-dimensionalized  observed length, $\hat{L}_{obs}$ after the contact:
\begin{eqnarray*}
\hat{L}_{obs}(\hat{t})=\frac{\hat{k}\hat{L}_c}{1+\hat{k}}+\frac{\hat{L}_c\hat{\gamma}_{el}\exp(\hat{\gamma}_{el}\hat{t})}{(1+\hat{k})\hat{\gamma}+ \hat{k}\hat{L}_c\exp(\hat{\gamma}_{el}\hat{t})}
     \label{elasticgrowth_solution},
\end{eqnarray*}
with:
$\hat{\gamma}_{el}=\hat{P}_c\left(1+(1-2\nu)\frac{\hat{k}}{1+\hat{k}}\right)-\hat{Y}_\sigma+\frac{ \hat{k}}{1+\hat{k}}$.
The solution saturates for a length increment after the contact:
\begin{eqnarray}
 \Delta\hat{L}_{obs}=\frac{\hat{P}_c-\hat{Y}_\sigma}{\hat{k}}.
\end{eqnarray}
The elastic growth model predicts a saturation of the observed (turgid) length which corresponds to a transient behavior of the physiological growth model more pronounced at high obstacle stiffness. In general both the relaxation rate ($\hat{\gamma}_i=\hat{k}/(1+\hat{k})+\hat{Y}_\sigma$) and the length increment differ with the elastic growth model outputs:
$$\Delta \hat{L}=\frac{\hat{k}+\hat{k}\hat{P_c}\left(1-2\nu\right)}{(1+\hat{k})\hat{P}_c+\hat{k}}\frac{\hat{P}_c-\hat{Y}_\sigma}{\hat{k}+(1+\hat{k})\hat{Y}_\sigma}.$$
As $\hat{P}_c\ll 1$, and  $\hat{Y}_\sigma\ll 1$, $\gamma_i$ equals $\gamma_{el}$ and $\Delta L$ equals $\Delta L_{obs}$  at the first order of approximation supposing $k\geqslant k_{cell}$; both models coincide in the early phase of the interaction (a few minutes for a $1\ cm$ long \textit{Chara corralina} internodal cell for an obstacle stiffness superior to $3142\ Nm^{-1} $). The morphoelastic model predicts the same asymptotic force independently of the obstacle stiffness (Figure \ref{Simu_analytique} (c)).
\section{Conclusion}\label{conclusion}
An apical compression is the simplest way to probe the mechanics of a cylindrical cell. The article derives the forces (before and after relaxation) exerted by the sensor on the compressed cell and it shows that the forces are independent on the pressure at the first order. The initial pressure, equilibrated by the tension in the cell wall, does not contribute to the forces. The forces only depend on the surface modulus $Eh$ multiplied  by the radius and the force drop observed during the relaxation is Poisson ratio dependent; the force drop is null when the cell wall is an incompressible material, that is for a Poisson ratio of 0.5, and is maximal for a zero Poisson ratio. A compression also induces a turgor pressure jump which drops strongly during relaxation due to water outflow: before the relaxation, the pressure jump is proportional to the surface modulus divided by the radius  while after the relaxation, the pressure jump is proportional to the initial pressure. According to the linear elasticity hypothesis $PR/Eh$ is inferior to $1\%$, meaning the pressure jump after relaxation can be neglected in front of the pressure jump before relaxation. As in the case of isotropic poroelastic gels, the ratio between the force before and the force after relaxation only depends on the Poisson ratio. However the ratio value for cylindrical cells $\left(\frac{5}{4}-\nu\right) /(1-\nu^2)$ is different from that for gels $ 2(1-\nu)$ \citep{cai2010poroelasticity}; interestingly in both cases the ratio equals one for an incompressible material. As a perspective, the model should be refined to describe the anisotropic properties of plant cell wall which could induce a stronger drop of the force or of the pressure during the relaxation.\\

The model explores how a stationary osmoregulation process coupled with the stress-formulated Lockhart's law predicts the dynamics of turgor pressure and cell length throughout the contact of a cell growing against an obstacle. The phenomenology of the growth against a stiff obstacle is reproduced: a transient growth arrest (Figure \ref{Simu_analytique}) followed by a second phase with a slower variation of growth velocity which remains strictly positive (See \citep{quiros2022plant} Figure 6 top left panel for forces superior to $0.04N$ ). Interestingly this regime corresponds to the growth behavior of \textit{Bacillus subtilis}, a Gramm positive bacteria, in gels with an agarose concentration ranging between $1\%$ and $8\%$ \citep{tuson2012measuring}. The article also provides the opportunity to compare a physiological model to a more phenomenological approach, coined morphoelasticity, which neglects the osmotic origin of the growth. Both models are calibrated to be equivalent before the contact with obstacle. The morphoelastic model and the physiological model are not equivalent in general; they only coincide for high obstacle stiffness ($k\geqslant k_{cell}$) on short time scale ($\gamma t\ll 1$). The physiological model could be extended to describe the invasive growth of fungi (hyphe) or of some plant organ (pollen tube, root hairs ...) \citep{sanati2013cellular} by adapting the framework to apical growth; the analytical approach is solvable as analytical solutions linking the force to the apical deformation are available \citep{couturier2022compression}.\\
A contact and more generally a force exerted on a cell is known to induce many biological regulations. Some regulations tend to stop growth (touch responses) \citep{braam2005touch}, \citep{coutand_biomechanical_2000}, \citep{bizet_3d_2016} while other regulations on the contrary tend to favor growth despite the cell wall tension drop, such as cell wall loosening \citep{green_metabolic_1971} or through the increase of osmotic pressure \citep{greacen_physics_1972}, \citep{atwell_physiological-responses_1988}). The deviations from the quantitative predictions of our model are a good base to quantify how these active responses act and differ from the stationary osmoregulation.\\

\textbf{Acknowledgement}. We thank Bruno Moulia for a stimulating discussion on stationary osmoregulation.

\textbf{Funding statement}. MBBT was supported by a grant overseen by the French National Research Agency (ANR) as part of the ‘Investissements
d’Avenir’ programme (ANR-11-LABX-0002-01, Lab of Excellence ARBRE). WK and EC were supported by ANR AnAdSpi ANR-20-CE30-0005-01.
\section{Annex}\label{Annex}
\subsection{Force and pressure before relaxation}\label{Force and pressure before relaxation}
We start with the equations of mechanical equilibrium before compression of the cell. The longitudinal $\epsilon_{LL,0}$ and circumferential $\epsilon_{\Theta\Theta,0}$ strains in the cell wall can be written as:
\begin{eqnarray}
\left(\begin{array}{c}
 \epsilon_{LL,0}  \\
\epsilon_{\Theta\Theta,0}
\end{array} \right)=\left(\begin{array}{c}
(1-2\nu)P_{0}R^*(1+\epsilon_{\Theta\Theta,0})/(2Eh) \\
(2-\nu)P_{0}R^*(1+\epsilon_{\Theta\Theta,0})/(2Eh)
\end{array} \right)
\label{eq_0}
\end{eqnarray}
with $\epsilon_{LL,0}=L_0/L^*-1$ and $\epsilon_{\Theta\Theta,0}=R_0/R^*-1$. At the first order in strain, (\ref{eq_0}) rewrites:
\begin{eqnarray}
\left(\begin{array}{c}
 \epsilon_{LL,0}  \\
\epsilon_{\Theta\Theta,0}
\end{array} \right)=\left(\begin{array}{c}
(1-2\nu)P_{0}R^*/(2Eh) \\
(2-\nu)P_{0}R^*(2Eh)
\end{array} \right)
\label{eq_0trunc}
\end{eqnarray}
leading to:
\begin{eqnarray}
R_0&=&R^*(1+(2-\nu)P_{0}R^*/(2Eh)),
\label{eq_R0}
\end{eqnarray}
\begin{eqnarray}
L_0&=&L^*(1+(1-2\nu)P_{0}R^*/(2Eh)).
\label{eq_L0}
\end{eqnarray}

Then we write the equations following the instantaneous compression before the relaxation occurs. The volume is conserved which implies:
\begin{eqnarray*}
R_{br}&=&R_0\sqrt{L_0/L_{ind}}.
\end{eqnarray*}
At first order,
\begin{eqnarray*}
R_{br}=R_0+R^*\frac{(L_0-L_{ind})}{2L^*}
\end{eqnarray*}
substituting with the values of (\ref{eq_R0},\ref{eq_L0}), it gives: 
\begin{eqnarray}
R_{br}=R^*\left(1+\left(2-\nu+\frac{1-2\nu}{2}\right)P_{0}R^*/(2Eh)-\frac{1}{2}(L_{ind}/L^*-1)\right)
\label{Rbr_inf}
\end{eqnarray}
Developing the formula (\ref{mechanica_equilibrium_br}) of the mechanical equilibrium, it gives:
\begin{eqnarray*}
\left(\begin{array}{c}
 L_{ind}/L^*-1  \\
 R_{br}/R^*-1
\end{array} \right)=\left(\begin{array}{c}
(1-2\nu)P_{br}R^*/(2Eh) -F_{br}/(2\pi E R^* h) \\
(2-\nu)P_{br}R^*/(2Eh)+\nu F_{br}/(2\pi E R^* h)
\end{array} \right)
\end{eqnarray*}
The force $F_{br}$ can be expressed: 
\begin{eqnarray*}
F_{br}&=&(\pi ER_{br}h) \frac{(-2+\nu)(L_{ind}/L^*-1)+(1-2\nu)(R_{br}/R^*-1)}{1-\nu^2},
\end{eqnarray*}
Substituting the expression for $R_{br}$ (\ref{Rbr_inf}) gives:
\begin{eqnarray*}
F_{br}&=&(\pi ER_{br}h) \frac{(-2+\nu)(L_{ind}/L^*-1)+(1-2\nu)\left(\left(2-\nu+\frac{1-2\nu}{2}\right)P_{0}R^*/(2Eh)-\frac{1}{2}(L_{ind}/L^*-1)\right)}{1-\nu^2}
\end{eqnarray*}
which rewrites neglecting second order terms: 
\begin{eqnarray*}
F_{br}&=&(\pi ER^*h) \frac{(-2+\nu)(L_{ind}/L^*-1)+(1-2\nu)\left(\left(2-\nu+\frac{1-2\nu}{2}\right)P_{0}R^*/(2Eh)-\frac{1}{2}(L_{ind}/L^*-1)\right)}{1-\nu^2}.
\end{eqnarray*}
It can be simplified by reintroducing $L_0$:
\begin{eqnarray}
F_{br}=(\pi ER^*h)\left(\frac{5}{2}-2\nu\right) \frac{(L_0-L_{ind})}{L^*(1-\nu^2)}
\label{F_br}
\end{eqnarray}
The mechanical equilibrium also provides the pressure:
\begin{eqnarray*}
P_{br}=(Eh)\frac{( R_{br}/R^*-1)+\nu ( L_{ind}/L^*-1)}{(1-\nu^2)R^*}
\end{eqnarray*}
Substituting the expression for $R_{br}$ and rearranging the term to make appear $L_0$ gives:
\begin{eqnarray}
P_{br}=P_0+Eh\frac{(1-2\nu)}{2(1-\nu^2)}\frac{L_0-L_{ind}}{L^*R^*}.
\label{P_br}
\end{eqnarray}

\subsection{Force and pressure after relaxation}\label{Force and pressure after relaxation}
The equation (\ref{mech_eq_osm_eq}) can be rewritten as:
\begin{eqnarray}
\left(\begin{array}{c}
\epsilon_{LL,ind}  \\
\epsilon_{\Theta\Theta,ar}
\end{array} \right)&=&\left(\begin{array}{c}
(1-2\nu)\mathcal{R} T N_i/(2E h\pi R^*(1+\epsilon_{\Theta\Theta,ar})L_{ind}) -F_{ar}/(2\pi E R^*(1+\epsilon_{\Theta\Theta,ar}) h) \\
(2-\nu)\mathcal{R} T N_i/(2E h\pi R^*(1+\epsilon_{\Theta\Theta,ar}) L_{ind}) +\nu F_{ar}/(2\pi E R^*(1+\epsilon_{\Theta\Theta,ar}) h)
\end{array} \right)
\label{eq_ar}
\end{eqnarray}
with $\epsilon_{LL,ind}= L_{ind}/L^*-1$ and $\epsilon_{\Theta\Theta,ar}= R_{ar}/R^*-1$. Multiplying the both sides of (\ref{eq_ar}) by $(1+\epsilon_{\Theta\Theta,ar})$ and truncating at the first order in strain gives:
\begin{eqnarray*}
\left(\begin{array}{c}
\epsilon_{LL,ind}  \\
\epsilon_{\Theta\Theta,ar}
\end{array} \right)&=&\left(\begin{array}{c}
(1-2\nu)\mathcal{R} T N_i/(2E h\pi R^*L_{ind}) -F_{ar}/(2\pi E R^* h) \\
(2-\nu)\mathcal{R} T N_i/(2E h\pi R^* L_{ind}) +\nu F_{ar}/(2\pi E R^* h)
\end{array} \right)
\end{eqnarray*}
which can be expanded in:
\begin{eqnarray}
\left(\begin{array}{c}
L_{ind}/L^*-1 \\
R_{ar}/R^*-1
\end{array} \right)&=&\left(\begin{array}{c}
(1-2\nu)\mathcal{R} T N_i/(2E h\pi R^*L_{ind}) -F_{ar}/(2\pi E R^* h) \\
(2-\nu)\mathcal{R} T N_i/(2E h\pi R^* L_{ind}) +\nu F_{ar}/(2\pi E R^* h)
\end{array} \right).
\label{eq_ar_trunc}
\end{eqnarray}
Substracting (\ref{eq_0trunc}) from (\ref{eq_ar_trunc}) gives:
\begin{eqnarray*}
\left(\begin{array}{c}
(L_{ind}-L_0)/L^* \\
(R_{ar}-R_0)/R^*-1
\end{array} \right)&=&\left(\begin{array}{c}
 -F_{ar}/(2\pi E R^* h) \\
\nu F_{ar}/(2\pi E R^* h)
\end{array} \right).
\end{eqnarray*}
which provides easily:
\begin{eqnarray}
R_{ar}&=&R_0+\nu R^* \frac{(L_0-L_{ind})}{L^*},
\label{Rar}
\end{eqnarray}
\begin{eqnarray}
F_{ar}&=&(2\pi ER^*h)
\frac{(L_0-L_{ind})}{L^*}.
\label{Far}
\end{eqnarray}
The apparent stiffness after relaxation $k_{ar}$ equals $(2\pi ER^*h)/L^*$.
The pressure after relaxation reads:
\begin{eqnarray*}
P_{ar}=\mathcal{R} T N_i/(\pi R_{ar}^2L_{ind})
\end{eqnarray*}
which at the first order in strain:
\begin{eqnarray*}
P_{ar}=P_0\left(1+(1-2\nu)\frac{(L_0-L_{ind})}{L^*}\right).
\label{Par}
\end{eqnarray*}

\subsection{First order solution for the osmotic equilibrium and the mechanical equilibrium with an obstacle}\label{osmotic equilibrium and mechanical}
Substituting (\ref{Pressure}) in (\ref{stress_axial_growth},\ref{stress_radial_growth}) gives:
\begin{equation}
\left(\begin{array}{c}
 L/L^*-1  \\
 R/R^*-1
\end{array} \right)=\left(\begin{array}{cc}
 1& -\nu  \\
-\nu& 1
\end{array} \right)\left(\begin{array}{c}
\mathcal{R} T N_i/(2Eh\pi R L)-k/(2\pi E R h)(L-L_c)_+   \\
\mathcal{R} T N_i/(Eh\pi R L)
\end{array} \right)
\label{mechanicaL_{obs}quilibrium_plus osmotic}
\end{equation}
Introducing the small deformations $\epsilon_{LL}$ and $\epsilon_{\Theta\Theta}$ gives:
\begin{equation}
\left(\begin{array}{c}
 \epsilon_{LL}  \\
\epsilon_{\Theta\Theta}
\end{array} \right)=\left(\begin{array}{cc}
 1& -\nu  \\
-\nu& 1
\end{array} \right)\left(\begin{array}{c}
\mathcal{R} T N_i/(2Eh\pi R^* L^*(1+\epsilon_{LL})(1+ \epsilon_{\Theta\Theta}))-k(L^*(1+\epsilon_{LL})-L_c)_+/(2\pi E R^*(1+\epsilon_{\Theta\Theta}) h)   \\
\mathcal{R} T N_i/(Eh\pi R^* L^*(1+\epsilon_{LL})(1+\epsilon_{\Theta\Theta}))
\end{array} \right)
\label{equilibre_epsilon}
\end{equation}
Multiplying both sides by $(1+\epsilon_{LL})(1+\epsilon_{\Theta\Theta})$ gives:
\begin{equation}
\left(\begin{array}{c}
 \epsilon_{LL} (1+\epsilon_{LL})(1+\epsilon_{\Theta\Theta}) \\
\epsilon_{\Theta\Theta}(1+\epsilon_{LL})(1+\epsilon_{\Theta\Theta})
\end{array}\right)= \left(\begin{array}{c}
(1-2\nu)\mathcal{R} T N_i/(2Eh\pi R^* L^*)-k(1+\epsilon_{LL})(L^*(1+\epsilon_{LL})-L_c)_+/(2\pi E R^* h)   \\
(2-\nu)\mathcal{R} T N_i/(2Eh\pi R^* L^*)+\nu k(1+\epsilon_{LL})(L^*(1+\epsilon_{LL})-L_c)_+/(2\pi E R^* h)
\end{array} \right)
\label{equilibre_epsilon_inter}
\end{equation}
At the first order  in $\epsilon_{LL}$ and $\epsilon_{\Theta\Theta}$:
\begin{equation}
\left(\begin{array}{c}
 \epsilon_{LL} \\
\epsilon_{\Theta\Theta}
\end{array}\right)= \left(\begin{array}{c}
(1-2\nu)\mathcal{R} T N_i/(2Eh\pi R^* L^*)-k(L^*(1+2\epsilon_{LL})-L_c(1+\epsilon_{LL}))_+/(2\pi E R^* h)   \\
(2-\nu)\mathcal{R}T N_i/(2Eh\pi R^* L^*)+\nu k(L^*(1+2\epsilon_{LL})-L_c(1+\epsilon_{LL}))_+/(2\pi E R^* h)
\end{array} \right)
\label{equilibre_epsilon_order_1}
\end{equation}
It rewrites:
$$\left(1+k(2L^*-L_c))/(2\pi E R^* h)\right)\epsilon_{LL}=(1-2\nu)\mathcal{R} T N_i/(2Eh\pi R^* L^*)-k(L^*-L_c)/(2\pi E R^* h) $$
which yields:
\begin{equation}
\epsilon_{LL}=\frac{(1-2\nu)\mathcal{R} T N_i/(2Eh\pi R^* L^*)-k(L^*-L_c)/(2\pi E R^* h)}{1+k(2L^*-L_c)/(2\pi E R^* h)}.
\label{First_order_strain}
\end{equation}
Exploiting:
\begin{equation}
\epsilon_{\Theta\Theta}+\nu \epsilon_{LL}=(2-2\nu^2)\mathcal{R} T N_i/(2Eh\pi R^* L^*),
\end{equation}
it provides $\epsilon_{\Theta\Theta}$:
\begin{equation}
\epsilon_{\Theta\Theta}=(1-\nu^2)\mathcal{R} T N_i/(Eh\pi R^* L^*)-\nu\epsilon_{LL}.
\end{equation}
The longitudinal stress reads:
\begin{equation}
\sigma_{LL}/E=(\epsilon_{LL}+\nu\epsilon_{\Theta\Theta})/(1-\nu^2)=\epsilon_{LL}+\nu\mathcal{R} T N_i/(Eh\pi R^* L^*).
\label{First_order_stress}
\end{equation}
\subsection{First order expression of the Lockhart's model}
Substituting (\ref{First_order_stress}) in the time derivative of the plasmolyzed length corresponding to the stress-based formulation Lockhart's law gives:
\begin{equation}
f_\sigma=m_\sigma E\left(\frac{(1-2\nu)\mathcal{R} T N_i/(2Eh\pi R^* )-kL^*(L^*-L_c)/(2\pi E R^* h)}{1+k(2L^*-L_c)/(2\pi E R^* h)}+\frac{\nu\mathcal{R} T N_i}{\pi E R^*h }-\frac{Y_\sigma L^*}{E}\right)_+.
\label{Lockhart_trunc}
\end{equation}
The growth ODE (\ref{Lockhart_stress}) thus reads:
\begin{equation*}
\frac{d L^*}{d t}=m_\sigma E\left(\frac{(1-2\nu)\mathcal{R} T N_i/(2Eh\pi R^* )-kL^*(L^*-L_c)/(2\pi E R^* h)}{1+k(2L^*-L_c)/(2\pi E R^* h)}+\frac{\nu\mathcal{R} T N_i}{\pi E R^*h }-\frac{Y_\sigma L^*}{E}\right)_+.
\end{equation*}
At $t=0$, the contact time, the force exerted by the obstacle is zero and  the classical formula (\ref{eq_L0}) for a turgid cylinder is applicable: $$L_c=L^*_c(1+(1-2\nu)\mathcal{R}TN_{i}(0))/(2\pi R^* hL_{c}^*E)).$$
Introducing the variables $\hat{P}_c=\mathcal{R}TN_{i}(0)/(2\pi R^* hL_{c}^*E)$,  $\hat{Y}_\sigma=Y_\sigma/E$, and substituting with the expression for $L_c$ provides:
\begin{equation}
\frac{d L^*}{d t}=m_\sigma E\left(\frac{(1-2\nu)\hat{P}_c\exp(\gamma t)L^*_c-kL^*(L^*-L^*_c(1+(1-2\nu)\hat{P}_c))/(2\pi E R^* h)}{1+k(2L^*-L^*_c(1+(1-2\nu)\hat{P}_c))/(2\pi E R^* h)}+2\nu \hat{P}_c\exp(\gamma t)L^*_c-\hat{Y}_\sigma L^*\right)_+.
\label{eqtLoc}
\end{equation}
The obstacle stiffness can be non-dimensionalized with $k_{ar}=2\pi R^* hE/L^*_c$, the cell stiffness after relaxation at contact (\ref{stiff_ar}), by introducing $\hat{k}=k/k_{ar}$:
\begin{equation}
\frac{d L^*}{d t}=m_\sigma E\left(\frac{(1-2\nu)\hat{P}_c\exp(\gamma t)L^*_c-\hat{k}L^*(L^*/L^*_c-(1+(1-2\nu)\hat{P}_c))}{1+\hat{k}(2L^*/L^*_c-(1+(1-2\nu)\hat{P}_c))}+2\nu \hat{P}_c\exp(\gamma t)L^*_c-\hat{Y}_\sigma L^*\right)_+.
\label{eqtLoc2}
\end{equation}
The growth ODE (\ref{eqtLoc2}) can be rewritten with the non-dimensional time and length variables $\hat{t}=m_\sigma E t$ and $\hat{L}^*=L^*/L_c^*$:
\begin{equation}
\frac{d \hat{L}^*}{d \hat{t}}=\left(\frac{(1-2\nu)\hat{P}_c\exp(\hat{\gamma} \hat{t})-\hat{k}\hat{L}^*(\hat{L}^*-(1+(1-2\nu)\hat{P}_c))}{1+\hat{k}(2L^*-(1+(1-2\nu)\hat{P}_c))}+2\nu \hat{P}_c\exp(\hat{\gamma} \hat{t})-\hat{Y}_\sigma \hat{L}^*\right)_+.
\label{eqtLoc3}
\end{equation}
with: $\hat{\gamma}=  \left(\hat{P_c}-\hat{Y}_\sigma\right)_+$.\\
It corresponds to a non-dimensional growth ODE function (\ref{Lockhart_stress_adim}):
\begin{equation}
\hat{f}_\sigma(\hat{L}^*,\hat{t})=\left(\frac{(1-2\nu)\hat{P}_c\exp(\hat{\gamma}\hat{t})-\hat{k}\hat{L}^*(\hat{L}^*-1-(1-2\nu)\hat{P}_c)}{1+\hat{k}(2\hat{L}^*-1-(1-2\nu)\hat{P}_c)}+2\nu\hat{P}_c\exp(\hat{\gamma}\hat{t})-\hat{Y}_\sigma\hat{L}^*\right)_+.
\end{equation}
The longitudinal elastic strain (\ref{First_order_strain}) can be rewritten in term of the non-dimensional variables:
\begin{equation}
\epsilon_{LL}=\frac{(1-2\nu)\hat{P}_c/\hat{L}^*\exp(\hat{\gamma}\hat{t})-\hat{k}(\hat{L}^*-1-(1-2\nu)\hat{P}_c)}{1+\hat{k}(2\hat{L}^*-1-(1-2\nu)\hat{P}_c)}.
\label{First_order_strain-st}
\end{equation}
The $\hat{L}^*$ derivative of $\hat{f}_\sigma$ reads:
\begin{equation}
\partial_{\hat{L}^*}\hat{f}_\sigma=-\left(\frac{\hat{k}\left(2\hat{L}^*-1-(1-2\nu)\hat{P}_c+2\hat{L}^*\epsilon_{LL}\right)}{1+\hat{k}\left(2\hat{L}^*-1-(1-2\nu)\hat{P}_c\right)}+\hat{Y}_\sigma \right).
\end{equation}
making use of (\ref{First_order_strain-st}) to introduce $\epsilon_{LL}$.\\
For $\hat{t}>0$ and a length which equals the contact length, the normalized length derivative reads:
\begin{equation}
\hat{f}_\sigma(\hat{L}^*=1,\hat{t})=\left(\frac{(1-2\nu)\hat{P}_c\exp(\hat{\gamma}\hat{t})+\hat{k}(1-2\nu)\hat{P}_c}{1+\hat{k}(1-(1-2\nu)\hat{P}_c)}+2\nu\hat{P}_c\exp(\hat{\gamma}\hat{t})-\hat{Y}_\sigma\right)_+,
\end{equation}
\begin{equation}
\partial_{\hat{L}^*}\hat{f}_\sigma(\hat{L}^*=1,\hat{t})=-\left(\frac{\hat{k}\left(1-(1-2\nu)\hat{P}_c+2\epsilon_{LL}\right)}{1+\hat{k}\left(1-(1-2\nu)\hat{P}_c\right)}+\hat{Y}_\sigma \right).
\end{equation}
Following the hypothesis of linear elasticity, longitudinal elastic strain  $\epsilon_{LL}=(1-2\nu)\hat{P}_c$ is small. Keeping the dominant order provides:
\begin{equation}
\hat{f}_\sigma(\hat{L}^*=1,\hat{t})=\left(\exp(\hat{\gamma}\hat{t})\hat{P}_c\left(\frac{1+2\nu\hat{k}}{1+\hat{k}}\right)+\hat{P}_c\left(\frac{(1-2\nu)\hat{k}}{1+\hat{k}}\right)-\hat{Y}_\sigma\right)_+,
\label{f_hat_first}
\end{equation}
\begin{equation}
\partial_{\hat{L}^*}\hat{f}_\sigma(\hat{L}^*=1,\hat{t})=-\left(\frac{\hat{k}}{1+\hat{k}}+\hat{Y}_\sigma \right).\label{df_hat_first}
\end{equation}

\subsection{Analytical solution of the linearized problem}\label{Analytical solution}
The equation (\ref{eqtLoc3}) linearized at $(\hat{L}^*=1,\hat{t})$ reads:
\begin{equation}
\frac{d \hat{L}^*}{d \hat{t}}\approx \hat{f}_\sigma\left(1,\hat{t}\right)+(\partial_{\hat{L}^*} \hat{f}_\sigma)\left(1,\hat{t}\right)\left(\hat{L}^*-1\right).
\label{lin_Lockhart_0}
\end{equation}
(\ref{f_hat_first}) tells $\partial_{\hat{L}^*} \hat{f}_\sigma$ is independent of the time at first order. The solution of (\ref{lin_Lockhart_0}) reads:
\begin{equation}
\hat{L}^*=1+\exp\left(\partial_{\hat{L}^*} \hat{f}_\sigma(1,0)\hat{t}\right)\int_{u=0}^{\hat{t}}\hat{f}_\sigma\left(1,u\right)\exp\left(-\partial_{\hat{L}^*} \hat{f}_\sigma(1,0)u\right)du.
\end{equation}
Once substituted with (\ref{lin_Lockhart_0}), the integral reads:
\begin{equation}
\int_{u=0}^{\hat{t}}\left(\exp\left((\hat{\gamma}+\hat{\gamma}_i)u\right)\hat{P}_c\left(\frac{1+2\nu\hat{k}}{1+\hat{k}}\right)+\exp\left(\hat{\gamma}_iu\right)\left(\hat{P}_c\left(\frac{(1-2\nu)\hat{k}}{1+\hat{k}}\right)-\hat{Y}_\sigma\right)\right)du
\end{equation}
with $\hat{\gamma}=\hat{P}_c-\hat{Y}_\sigma$ and $\hat{\gamma}_i=\frac{\hat{k}}{1+\hat{k}}+\hat{Y}_\sigma$.
The solution can be explicited:
\begin{equation}
\hat{L}^*(\hat{t})\approx 1+\hat{\beta}\left(\exp\left(\hat{\gamma} \hat{t}\right)-1\right)+(1-\hat{\beta})\frac{\hat{\gamma}}{\hat{\gamma}_i}\left(1-\exp\left(-\hat{\gamma}_i \hat{t}\right)\right),
\label{Analyticd}
\end{equation}
with:
\begin{equation*}
\hat{\beta}=\frac{\hat{P_c}\left(1+2\nu\hat{k}\right)}{(1+\hat{k})\hat{P_c}+\hat{k}}.
\end{equation*}
For a $\textit{Chara corallina}$ internode of length at contact $L_c=1cm$, the mechanical parameters are of the following order of magnitude  \citep{huang2012modelling}:
\begin{equation*}
E=1GPa,\ h=10\mu m,\ \nu=0.3,\ R^*=500\mu m.
\end{equation*}
The Lockhart's parameters expressed in pressure read:
\begin{equation*}
P=0.55MPa,\ Y_P=0.5MPa,\ m_P=2.6000\times 10^{-5} (MPa)^{-1} s^{-1},
\end{equation*}
which correpond to Lockhart's parameters expressed in stresses obtained with these formula:
\begin{equation*}
\sigma_{LL}=\frac{PR}{2h},\ Y_\sigma=\frac{Y_PR}{2h},\ m_\sigma=\frac{2 hm_P}{R}.
\end{equation*}
It corresponds to a cell stiffness after relaxation: $k_{ar}=\pi\times 10^3N/m$.

The hypothesis of our model do not apply to a maiz root growth zone. Nonetheless  we provide its parameters in order to provide a qualitative comparison point with \textit{Chara corallina}. Approximating the maize root growth zone by a single cell of contact length $L_c=1cm$,  the mechanical parameters of the root taken as a plain rod are the following (Paragraph 4.2.6.3 of \citep{quiros2023biomecanique} ):
\begin{equation*}
E_{rod}=24MPa,\ R^*=450\mu m.
\end{equation*}
It corresponds to a stiffness
\begin{equation*}
k_{cell}=\frac{\pi (R^*)^2 E_{rod}}{L_c}=1527N/m
\end{equation*}
The Lockhart's parameters expressed in pressure read \citep{frensch_rapid_1995}:
\begin{equation*}
P=0.7MPa,\ Y_P=0.6MPa,\ m_P=8\times 10^{-4} (MPa)^{-1} s^{-1}.
\end{equation*}
\begin{figure}[htbp]
\begin{center}
\includegraphics[scale=0.8]{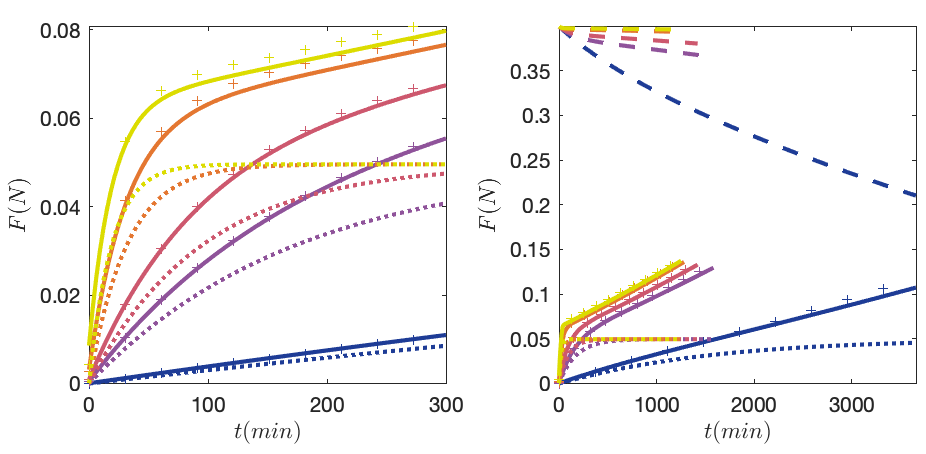}
\caption{Force versus time in real units. $k=k_{cell}/100$ (blue), $k=k_{cell}/10$ (violet), $k=k_{cell}/5$ (dark red), $k=k_{cell}$ (orange), $k=5k_{cell}$ (yellow). Other parameters are detailed in the annex (\ref{Analytical solution}). Plain lines and $+$ symbols stand for the numerical and analytical solutions respectively. The dotted line is the morphoelastic solution. The dashed line is the Euler criterion corresponding to the numerical solution. The left plot is a zoom on the five first hours of growth (corresponding to the duration of the experiment of reference \citep{proseus2000turgor}, Figure 5); the right plot is integrated until one of the strains equals $3\%$.}
\label{Simu_analytique_Euler}
\end{center} 
\end{figure}

\subsection{Asymptotic behavior of the Lockhart's model with an obstacle}\label{Asymptotic behavior}
The time derivative of (\ref{Lockhart_stress}) gives:
\begin{equation*}
\frac{d^2 L^*}{d t^2}=\partial_t f_\sigma(L^*,t)+\partial_L^* f_\sigma(L^*,t)\frac{d L^*}{d t}.
\end{equation*}
Let suppose $\frac{d L^*}{d t}$ equals zero for a given $t_{zero}$:
\begin{equation*}
\frac{d^2 L^*}{d t^2}(t_{zero})=\partial_t f_\sigma(L^*(t_{zero}),t_{zero})
\end{equation*}
The time derivative reads:
\begin{equation*}
\partial_t f_\sigma(L^*(t_{zero}),t_{zero})=m_\sigma \gamma \frac{\mathcal{R} T N_i(t_{zero})}{2h\pi R^* L^*} \left(\frac{1+2\nu k(2L^*-L_c)/(2\pi ER^* h)}{1+k(2L^*-L_c)/(2\pi ER^* h)}\right),
\end{equation*}
which implies:
\begin{equation*}
\frac{d^2 L^*}{d t^2}(t_{zero})>0.
\end{equation*}
$L^*$ is thus strictly increasing with no fixed point; $L^*$ asymptotic limit reads:
\begin{equation*}
\lim_{t\to \infty}L^*=+\infty.
\end{equation*}
The longitudinal stress reads:
\begin{equation*}
\sigma_{LL}= \left(\frac{(1-2\nu)\mathcal{R} T N_i/(2Eh\pi R^* L^*)-k(L^*-L_c)/(2\pi R^* h)}{1+k(2L^*-L_c)/(2\pi E R^* h)}+\frac{\nu\mathcal{R} T N_i}{\pi R^*h L^*}\right).
\end{equation*}
When $t\to \infty$ it can be rewritten:
\begin{equation*}
\sigma_{LL}= -\frac{E}{2}+\frac{\nu\mathcal{R} T N_i}{\pi E R^*h L^*}+o_{t\to\infty}\left(\frac{\mathcal{R} T N_i}{\pi  R^*h L^*}\right).
\end{equation*}
(\ref{Lockhart_trunc}) can be rewritten:
\begin{equation}
\frac{d L^*}{d t}=m_\sigma\left( -\frac{EL^*}{2}+\frac{\nu\mathcal{R} T N_i}{\pi R^*h }-Y_\sigma L^*+o_{t\to\infty}\left(\frac{\mathcal{R} T N_i}{\pi R^*h}\right)\right)_+.
\label{eq_asymp}
\end{equation}
Let write the time derivative of $\rho=L^*/N_i$:
\begin{equation*}
\frac{d \rho}{d t}=\frac{1}{N_i}\frac{dL^*}{dt}-\frac{L^*}{N_i^2}\frac{dN_i}{dt}
\end{equation*}
Substituting the value of $N_i$ and of $\dot{L^*}$:
\begin{equation*}
\frac{d \rho}{d t}=m_\sigma\left( -\frac{E}{2}\rho+\frac{\nu\mathcal{R} T}{\pi R^*h }-Y_\sigma \rho+o_{t\to\infty}\left(\frac{\mathcal{R} T }{\pi R^*h}\right)\right)_+-\gamma\rho
\end{equation*}
Substituting the value of $\gamma=m\sigma(P_0R^*/(2h)-Y_\sigma)$ gives:
\begin{equation}
\frac{d \rho}{d t}=m_\sigma\left( -\frac{E}{2}-P_0R^*/(2h)\right)\rho+m_\sigma\frac{\nu\mathcal{R} T}{\pi R^*h }+o_{t\to\infty}\left(m_\sigma\frac{\mathcal{R} T }{\pi R^*h}\right)
\label{ode_eq}
\end{equation}
Let take a very high $t_o$, and integrate the ODE from the time $t_o$ (\ref{ode_eq}): 
\begin{equation}
\rho=\exp\left(m_\sigma\left( -\frac{E}{2}-P_0R^*/(2h)\right)(t-t_o)\right)\left(\rho_0-\frac{2\nu\mathcal{R} T}{\pi R^*h(Eh+P_0R^*)}\right)+\frac{2\nu\mathcal{R} T}{\pi R^*h(Eh+P_0R^*)}
\end{equation}
Finally, $L^*$ has a simple equivalent for high $t$:
\begin{equation}
L^*\sim_{t\to\infty} \frac{2\nu \mathcal{R} T N_i(t)}{\pi R^*(P_0R^*+Eh)},
\end{equation}
it corresponds to a value of elastic strain outside of the range of linear elasticity:
\begin{equation}
\lim_{t\to\infty}\epsilon_{LL}= -1/2,
\end{equation}
and an asymptotic pressure:
\begin{equation}
\lim_{t\to\infty}P= \frac{P_0+Eh/R^*}{2}.
\end{equation}

\subsection{Cell buckling and barreling}\label{non_linearities}
The Euler criterion that gives the force threshold for buckling of an hollow cylinder is provided by \citep{landau1959theory}:
\begin{eqnarray}
F_b=\frac{\pi^2(R^*)^3h E}{(L^*)^2}
\end{eqnarray}
The force exerted by the obstacle remains well under the Euler criterion during the whole growth (Figure \ref{Simu_analytique_Euler}, Right Panel).\\

The barreling criterium can be calculated by a complex numerical procedure described in \citep{goriely2008nonlinear} but not repeated herein. In \citep{goriely2008nonlinear} Figure 3 for a slightly different constitutive law (neo-Hookean $C_1=1$) and a slightly different geometry ($h/R=0.01$ instead of $h/R=0.02$ for \textit{Chara corallina}),  the barreling criterium expressed lays under the Euler criterium for the ratio ($R/L=0.05$) of \textit{Chara corallina} meaning the force for barreling is higher than the Euler criterion; this result should remain valid for a $1\ cm$ internode.

\subsection{Elastic growth model}\label{Elastic growth model}
The contact scenario is modeled by one spring of a constant stiffness $k_{cell}$ and an increasing target length $L_{tar}$ and one spring of constant rest length $L_c$ and a constant stiffness $k$ (See Figure \ref{Experiences}c). The observed length $L_{obs}$ is provided by the mechanical equilibrium:
\begin{equation}
    k_{cell}(L_{obs}-L_{tar})+k(L_{obs}-L_c)=0,
\end{equation}
thus:
\begin{equation}
    L_{obs}=\frac{kL_c+k_{cell}L_{tar}}{k_{cell}+k}.
    \label{L_obslastic_growth}
\end{equation}

To retrieve the phenomenology, we suppose the target length evolution rate decreases linearly with the force:
\begin{equation}
    \dot{L}_{tar}=L_{tar}f(L_{tar})
     \label{Target_growth_law}
\end{equation}
with $f(L_{tar})= (c_1-c_2 k(L_{obs}-L_c))$ the growth rate. 
$c_1$ corresponds to the growth rate before the contact:
\begin{equation}
c_1=m_\sigma\left(\frac{P_c R}{2h}-Y_\sigma\right).
\end{equation}
$c_2$ can be estimated by considering the growth rate tangent at the contact time:
\begin{equation}
c_2=\frac{(\frac{\dot{L}}{L}-c_1)}{F}.
\end{equation}
Approximating $dL/dt/L$ by $ dL^*/dt/L^*$ at a contact time which is reasonable according to the analytical solution for the Lockhart model:
\begin{equation}
c_2=\frac{m_\sigma}{2\pi R h}.
\end{equation}
Substituting  (\ref{L_obslastic_growth}) in (\ref{Target_growth_law}) gives:
\begin{equation}
 \frac{d L_{tar}}{d t}   =L_{tar}\left(m_\sigma\left(\frac{P_c R}{2h}-Y_\sigma\right)- \frac{m_\sigma}{2\pi R h}\frac{ k_{cell}k}{k_{cell}+k}(L_{tar}-L_c)\right)_+.
     \label{Target_growth_equation}
\end{equation}
The equation can be non-dimensionalized using the following non-dimensional variables:
\begin{equation}
\hat{t}=m_\sigma E t,\ \hat{L}_{tar}=\frac{L_{tar}}{L_c^*},\ \hat{P}_c=\frac{\mathcal{R}TN_{i}(0)}{2\pi R^* hL_{c}^*E},\ \hat{Y}_\sigma=\frac{Y_\sigma}{E},
\label{nondim3}
\end{equation}
\begin{equation}
\hat{k}=\frac{k}{k_{cell}},\ \hat{\gamma}=  \left(\hat{P_c}-\hat{Y}_\sigma\right)_+,\ \hat{\gamma}_{el}=\hat{\gamma}+\frac{ \hat{k}}{1+\hat{k}}\hat{L}_c.
\label{nondim4}
\end{equation}
The non-dimensionalized equation reads:
\begin{equation}
 \frac{d \hat{L}_{tar}}{d\hat{t}}=\hat{L}_{tar}\left(\hat{\gamma}_{el}- \frac{ \hat{k}}{1+\hat{k}}\hat{L}_{tar}\right)_+.
     \label{Target_growth_equation_adim}
\end{equation}
It rewrites:
\begin{equation*}
 \frac{d \hat{L}_{tar}}{d\hat{t}}\left(\frac{1}{\hat{L}_{tar}}+\frac{\frac{ \hat{k}}{1+\hat{k}}}{\hat{\gamma}_{el}- \frac{ \hat{k}}{1+\hat{k}}\hat{L}_{tar}}\right)=\hat{\gamma}_{el}.
\end{equation*}
We suppose the contact takes place at time $t=0$; it implies $L_{tar}(0)=L_c$.
The expression (\ref{Target_growth_equation_adim}) can be integrated:
\begin{eqnarray*}
    \log(\hat{L}_{tar}/\hat{L}_c)-\log\left(\frac{\hat{\gamma}_{el}- \frac{ \hat{k}}{1+\hat{k}}\hat{L}_{tar}}{\hat{\gamma}}\right)
 =\hat{\gamma}_{el}\hat{t}.
     \label{Target_growth_integration}
\end{eqnarray*}
Finally:
\begin{eqnarray}
 \hat{L}_{tar}(\hat{t})=\frac{\hat{L}_c\hat{\gamma}_{el}\exp(\hat{\gamma}_{el}\hat{t})}{\hat{\gamma}+\frac{ \hat{k}\hat{L}_c}{1+\hat{k}}\exp(\hat{\gamma}_{el}\hat{t})}.
     \label{Target_growth_solutiona}
\end{eqnarray}
The solution saturates for an increment:
\begin{eqnarray*}
 \Delta \hat{L}_{tar}=\frac{\hat{\gamma}_{el}(1+\hat{k})}{\hat{k}}-\hat{L}_c.
\end{eqnarray*}
Substituting with the value of $\gamma$
and $ \Delta \hat{L}_{tar}$ rewrites:
\begin{eqnarray*}
 \Delta \hat{L}_{tar}=\frac{\hat{\gamma}(1+\hat{k})}{\hat{k}}.
\end{eqnarray*}
Combining     (\ref{Target_growth_solutiona}, \ref{L_obslastic_growth}) provides the observed length dynamics:
\begin{eqnarray}
\hat{L}_{obs}(\hat{t})=\frac{\hat{k}\hat{L}_c}{1+\hat{k}}+\frac{\hat{L}_c\hat{\gamma}_{el}\exp(\hat{\gamma}_{el}\hat{t})}{(1+\hat{k})\hat{\gamma}+ \hat{k}\hat{L}_c\exp(\hat{\gamma}_{el}\hat{t})}
     \label{elasticgrowth_solutiona},
\end{eqnarray}
which saturates for a non-dimensionalized observed length increment:
\begin{eqnarray}
 \Delta\hat{L}_{obs}=\frac{\hat{\gamma}}{\hat{k}}.
\end{eqnarray}

\bibliographystyle{vancouver}

\end{document}